\documentstyle[12pt]{article}
\newcommand{\be}{\begin{equation}}
\newcommand{\ee}{\end{equation}}
\begin{document}
\def\theequation{\arabic{section}.\arabic{equation}}
\begin{titlepage}
\title{A crucial ingredient of inflation}
\author{Valerio Faraoni$^{1,2}$ \\ \\
{\small \it $^1$ Research Group in General Relativity (RggR)} \\ 
{\small \it Universit\'e Libre de Bruxelles,  Campus Plaine CP 231} \\
{\small \it  Blvd. du Triomphe, 1050 Bruxelles, Belgium}\\ \\
{\small \it $^2$ INFN-Laboratori Nazionali di Frascati, 
P.O. Box 13, I-00044 Frascati, Roma (Italy)}
}
\date{} 
\maketitle   \thispagestyle{empty}  \vspace*{1truecm}
\begin{abstract} 
Nonminimal coupling of the inflaton field to the Ricci
curvature of spacetime is generally unavoidable, and the paradigm of
inflation should be generalized  by including the corresponding term $\xi
R \phi^2/2$ in the Lagrangian of the inflationary theory. This paper
reports on the status  of the programme of generalizing inflation. 
First, the problem of finding the correct value (or set of values) of the
coupling constant $\xi$ is
analyzed; the result has important consequences for the success or failure
of inflationary scenarios. Then, the slow-roll approximation to
generalized inflation is studied. Both the unperturbed inflating universe
models and scalar/tensor perturbations are discussed, and open problems
are pointed out.
\end{abstract}
\vspace*{1truecm} 
\begin{center} 
Talk given at {\em Recent Developments in Gravitation, Cosmology and
Quantum Field Theory}, Peyresq, France (June 2000).
\end{center}     
\end{titlepage}   

\section{Introduction}

\setcounter{equation}{0}

Cosmic inflation is  a period of accelerated expansion of the universe
during its early phase: provided that inflation proceeds for a
sufficiently long time (such that the cosmic expansion in the inflationary
period is about $60$ e-folds) and that physical criteria for a
successful description of the universe \cite{KT,Linde} are met, inflation
solves
the classic problems of the standard big bang cosmology (the horizon,
flatness, and monopole problem \cite{KT,Linde}).  In addition, inflation
provides, as a bonus, a mechanism  (quantum fluctuations of the inflaton
field) to generate density perturbations, the seeds of structures
 observed in the universe today (galaxies, clusters, and superclusters). 

Nowadays, this last aspect is regarded as the main motivation to pursue
research on inflation (e.g. \cite{Liddlereview}). There
are many
scenarios of inflation, but no ``standard model'' is universally
accepted: inflation has been called a ``paradigm in search of a model''.  
In the vast majority of inflationary scenarios, the cosmic
acceleration is
driven by one (or more) scalar field(s): although there are exceptions
(e.g. the scenario of Starobinsky \cite{Starobinsky}), a scalar field is
sometimes added even to these scenarios in order to ``help'' inflation 
\cite{Maeda89PRD}.

The inflaton field $\phi$ satisfies the Klein-Gordon equation: for reasons
explained below, when generalizing the latter from Minkowski space to a
curved space, one needs to introduce, in general, a nonminimal coupling
term between the scalar $\phi$ and the Ricci curvature of spacetime $R$ as
follows:
\be  \label{KG}
\Box \phi -\frac{dV}{d\phi} -\xi R \phi =0 \; ,
\ee
where $\Box=g^{\mu\nu} \nabla_{\mu}\nabla_{\nu}$ is d'Alembert's operator
on a curved space,   $V ( \phi )$ is the scalar field self-interaction 
potential, and $\xi$ is a dimensionless coupling constant. The classic
works on inflation neglected the $-\xi R \phi$ term in Eq.~(\ref{KG})
(which is equivalent to assume that $\xi=0$); hereafter, this theory is
called {\em ordinary inflation}, as opposed to {\em generalized
inflation}, which corresponds to $\xi \neq 0$. As explained in the next
section, almost always the introduction of a nonminimal (i.e. $\xi \neq
0$ in Eq.~(\ref{KG})) coupling is not an option; rather, it is
unavoidable.
This fact is not well known to cosmologists and has profound consequences
for the physics of the inflaton. Then, given the unavoidability of
nonminimal coupling (hereafter NMC), one needs to rethink inflation by
appropriately
including  terms corresponding to $\xi \neq 0$ in the relevant equations.
This was already done for {\em specific} inflationary scenarios by a
number of authors
\cite{NMCinflation,SalopekBondBardeen89,FakirUnruh,FakirUnruh90a,
FakirUnruh92apj,HwangNoh98};
however, the
approach adopted
was largely one in which the coupling constant $\xi$ is regarded as an
extra parameter of inflation that can be used at one's will in order to
cure pre-existing problems of the inflationary scenario. To make an
example, chaotic inflation with quartic self-interaction $V( \phi
)=\lambda \phi^4$ and $\xi=0$ is fine-tuned: the amplitude of
anisotropies of the cosmic microwave background requires $\lambda \leq
10^{-12}$, a figure that makes the scenario uninteresting from the point
of view of particle physics which originally motivated it. The 
fine-tuning is significantly reduced if one introduces nonminimal
coupling with $\xi <0$ and $\left| \xi \right| \simeq 10^4$
\cite{MakinoSasaki91,FakirUnruh}; the price to pay for reducing the
fine-tuning of $\lambda$ is
the fine-tuning of $\xi$. We disagree with the philosophy of this
approach because the
coupling constant $\xi$ has, in general, a well-defined value in
nature\footnote{Exceptions, discussed later, are the cases in which $\xi$
is a running coupling in GUT theories, or when first loop corrections
are taken into account.} 
and is not an extra free parameter of the theory. In Sec.~3 we review the
known prescriptions for the value of the coupling constant $\xi$ and make
clear that, not only $\xi \neq 0$ in the general case, but also that 
fine-tuning $\xi$ is not a possibility. We then proceed to analyze the
consequences of including NMC into the equations of inflation. The study
necessarily proceeds at two levels: first, one has to consider the {\em
unperturbed} background universe; and then one continues with the study of
scalar and tensor {\em perturbations} of the fixed inflationary background
universe. The amplitudes and spectra of perturbations are very 
important since they leave a detectable imprint in the cosmic microwave
background.

Temperature anisotropies in the sky, likely the fingerprints
of inflation, have been discovered by the {\em COBE} satellite
\cite{smootetal} and their
experimental study is one of the primary goals of current cosmology. Major
improvements will come with the {\em MAP} \cite{MAP} and {\em PLANCK}
\cite{PLANCK} satellites to be launched, respectively, in the years 2001
and 2007.

In this paper we approach the task of reformulating generalized inflation
(i.e. including the $\xi \neq 0$ terms in the picture) from  a {\em
general} point of view, i.e. we do not adopt a specific inflationary
scenario. The results for the unperturbed universe are presented in
Secs.~3 and~4.  

A special role is played by the slow-roll approximation: apart from two
exceptions (power-law inflation and the string-inspired,
toy model of Ref.~\cite{Easther}-see also Ref.~\cite{MartinSchwarz}), one
cannot exactly solve the
equations of inflation (both unperturbed and perturbed), and one needs to
resort to the slow-roll approximation. The latter has been discussed in
great detail for minimal (i.e. $\xi=0$) coupling (see
Ref.~\cite{Lidseyetal} for a recent review), and is much needed also in
the case of 
nonminimal coupling, for which the equations are even more complicated.
Slow-roll generalized inflation is discussed in Sec.~4. The study of
scalar
and tensor perturbations with nonminimal coupling is the subject of
Sec.~5, where previous results are reviewed and completed to obtain
explicit formulas for the observables of inflation.
de Sitter solutions play the role of attractors of inflationary solutions
in the phase space for generalized, as well as ordinary, inflation; it 
is this fact that ultimately gives  meaning to the slow-roll
approximation. Sec.~6 contains a list of open problems and 
the conclusions.

\section{Nonminimal coupling of the scalar field}

\setcounter{equation}{0}

The action of gravity plus a nonminimally coupled scalar field as matter
is 
\be  \label{action}
S=\int d^4x \sqrt{-g} \left[ \frac{R}{2\kappa} -\frac{1}{2} \nabla^c
\phi \nabla_c \phi -V( \phi ) -\frac{\xi}{2}R\phi^2 \right] \; , 
\ee
where $\kappa\equiv 8\pi G$, $G$ is Newton's constant and, apart from
minor differences, we adopt the
notations and conventions\footnote{The metric signature 
is --~+~+~+ and $G$ denotes Newton's constant. The speed of light and
Planck's constant assume the value
unity and $m_{pl}=G^{-1/2}$ is the Planck mass.  The components of the
Ricci tensor are given in terms of the
Christoffel symbols
$\Gamma_{\alpha\beta}^{\delta}$ 
by $R_{\mu\rho}=
\Gamma^{\nu}_{\mu\rho ,\nu}-\Gamma^{\nu}_{\nu\rho ,\mu}+
\Gamma^{\alpha}_{\mu\rho}\Gamma^{\nu}_{\alpha\nu}-
\Gamma^{\alpha}_{\nu\rho}\Gamma^{\nu}_{\alpha\mu} $, and  
$\Box \equiv g^{ab}\nabla_{a}\nabla_{b}$.}
of Ref.~\cite{Wald}.

\subsection{Why $\xi\neq 0$~?}

The nonminimal coupling of the scalar $\phi$ described by Eq.~(\ref{KG})
was apparently introduced for the first time by Chernikov and Tagirov
\cite{ChernikovTagirov}, although it is better known from the
work of Callan, Coleman and Jackiw \cite{CCJ}. Why should one consider
$\xi \neq 0$~? The answers are numerous: a nonzero $\xi$ is generated by
quantum corrections even if it is absent in the classical action
\cite{BirrellDavies,BirrellDavies80,NelsonPanangaden82,FordToms82,
ParkerToms85,Ford87}.
If
one prepares a classical theory with $\xi=0$,
renormalization shifts it to one with $\xi \neq 0$. Even though the
shift is small, it can have a tremendous effect on an inflationary
scenario. The classical example of this effect is related to chaotic
inflation \cite{FutamaseMaeda89}: the shift $\xi=0 \rightarrow
\xi_{renormalized}=10^{-1}$ (a typical value predicted by renormalization
\cite{AllenIshihara}) is sufficient to ruin the chaotic inflationary
scenario with potential $V=\lambda \phi^4$
\cite{FutamaseMaeda89,FutRotMat89}. 

Another reason  to include a $\xi \neq 0$ term in the coupled
Einstein-Klein-Gordon equations is that it is required by
renormalization of the theory (this was the motivation for the
introduction of NMC by Callan, Coleman and Jackiw \cite{CCJ}). It has also
been argued (see below) that a NMC term is expected at high curvatures
\cite{FordToms82,Ford87},
and that classicalization of the universe in quantum cosmology requires
$\xi \neq 0$ \cite{Okamura}.  A
pragmatic point of view  would
be that, since NMC may be crucial for the success or failure of
inflation
\cite{Abbott81,FutamaseMaeda89,FutRotMat89,FaraoniPRD96,Calgary}, one
better take it into account and decide {\em a posteriori} whether its
presence is negligible or not.

In relativity, it turns out that any value of $\xi$ different from $1/6$
(``conformal coupling'', the value that makes the Klein-Gordon
equation (\ref{KG}), and the physics of $\phi$, conformally invariant if
$V=0$ or $V=\lambda \phi^4$ \cite{Wald}) spoils the Einstein equivalence
principle and is therefore not admissible
\cite{SFFS,GribPoberii}.

Whichever point of view one adopts, with motivations arising in a range of
areas as wide as quantum
field theory in curved spaces, wormholes \cite{wormholes}, black holes 
\cite{blackholes}, boson stars \cite{Jetzer,VanderBij}, specific
inflationary scenarios, a pure
relativist's approach, or merely a pragmatic one, the message is that in
general it is wise not to ignore the NMC term by setting $\xi=0$, as done
in ordinary inflation. Although the inclusion of NMC makes the analysis
considerably more difficult, and it was unknown in the early, pioneering
days  of inflationary theory, the times are mature for the inclusion of
NMC in the theory.

\subsection{What is the value of $\xi$~?}

It is plausible that the value of the coupling constant $\xi$ be fixed by
the physics of the problem, and not be left to the choice of the
theoretician as a free parameter.  A 
particle physicist's answer to the question ``what is the value of
$\xi$~?''
differs according to the theory of the scalar field employed.

If $\phi$ is a
Goldstone boson in a theory with a spontaneously broken global symmetry, 
then $\xi=0$ \cite{VoloshinDolgov}. If the scalar field
$\phi$ is associated to a composite particle, the value of $\xi$ is 
fixed by the dynamics of its components. In particular, in the
large $N$ approximation to the Nambu-Jona-Lasinio model, the value
$\xi=1/6$ 
was found \cite{HillSalopek92}. In the 
$O(N)$-symmetric model with quartic self interaction, in which the
constituents of the $\phi$-particle are  themselves bosons, 
$\xi$
depends on the coupling constants $\xi_i$ of the elementary scalars
\cite{Reuter94}. 
In Einstein's gravity with the potential 
\begin{equation}
 V( \phi)=V_0+\frac{m^2}{2}\, \phi^2+\frac{\eta}{3!}\,\phi^3
+\frac{\lambda}{4!}\,\phi^4 
\end{equation}
and back-reaction, the value $\xi=0$ was found
\cite{Hosotani85,ParkerToms}. 
Higgs fields in the standard model have values of  
$\xi$ in the range $\xi \leq 0$, $\xi \geq 1/6$ \cite{Hosotani85}.

A great deal of results is available in the renormalization group approach
to quantum field theory in curved spaces. It is shown in
Refs.~\cite{Buchbinderetal}  that in asymptotically free GUTs, depending
on
the gauge group
employed ($SU(2)$, $SU(N)$, $SO(N)$,~...) and on the matter content, $\xi
$ is a running coupling that converges to $1/6$ (asymptotic conformal
invariance), or to a value $\xi_0$ determined by the initial conditions
(usually this occurs for supersymmetric GUTs) or (formally), $\left| \xi
( \tau ) \right| \rightarrow +\infty$. The last behaviour is often
characteristic
of large gauge groups ($SU(10)$, $SO(10)$,~...). Here $\tau$ is a
renormalization group parameter, with $\tau \rightarrow +\infty$
corresponding to strong curvature and early universe situations. In
Ref.~\cite{book} it was shown that also in asymptotically free GUTs
with $SU(5)$ as the gauge group, $\left| \xi ( \tau ) \right| \rightarrow
+ \infty$. Finite GUTs (another class of GUTs) behave similarly to
asymptotically free GUTs, with $\xi (\tau ) $ converging to $1/6$, or to 
an
initial value $\xi_0$ (e.g. for $N=4$ supersymmetry), or to infinity.
Moreover, for
finite GUTs the convergence of $\xi ( \tau )$ to its asymptotic value as
$\tau \rightarrow +\infty$ is much faster than in asymptotically free GUT
models (indeed, the convergence is exponentially fast \cite{BOL89,book}).
Hence, the asymptotic value of $\xi$ in the early universe strongly
depends on the choice of the specific GUT and of its gauge group and
matter
content.

The problem of the value of $\xi$ in this context is not an easy one, as
is clear from the example case of the simple $\lambda \phi^4 $ theory. The
latter is asymptotically free in the infrared limit ($\tau
\rightarrow - \infty$), which does not correspond to high curvature.
Nevertheless, it was shown in Ref.~\cite{Buchbinderetal} that $\xi( \tau )
\rightarrow 1/6$ as $\tau \rightarrow -\infty$. In the limit $\tau
\rightarrow +\infty$ of strong curvatures, one cannot answer the question
of the asymptotic value of $\xi( \tau ) $ since the theory is
contraddictory
(not asymptotically free) in this limit. Nevertheless, an {\em exact}
renormalization group approach to the $\lambda \phi^4$ theory shows that
$\xi=1/6$ is indeed a stable fixed point of the exact renormalization
group \cite{ParkerToms}. 

So far, controversies on these results only arose  for a restricted class
of specific models \cite{Bonanno}. The divergence of the running
coupling $\xi$ as the energy scale and the curvature and temperature
increase going back in time in the history of the universe, has been
introduced in cosmology \cite{HillSalopek92} and exploited to make the
chaotic inflationary scenario with $\xi <0$ more plausible in the cases in
which $\left| \xi ( \tau ) \right| \rightarrow +\infty$
\cite{FutamaseTanaka99}. The divergence of the coupling $\xi$ is also
crucial for the success of the so-called ``geometric reheating'' of the
universe after inflation \cite{BassettLiberati}, in which particles are
created due to the strong coupling of the inflaton to the Ricci curvature
$R$, instead of the usual coupling of $\phi$ to other fields.

First loop corrections to the classical theory make $\xi$ likely to be
a running parameter which depends on the Ricci curvature: in 
Refs.~\cite{FordToms82,ParkerToms} the effective coupling
\be
\xi_{eff}=\xi  +\frac{\lambda}{16\pi^2}\left( \xi-\frac{1}{6}\right) \ln s
\;,
\ee
was found for the self-interaction potential $\lambda \phi^4 /4! $,
where $s$ is
a parameter that tends to zero in the renormalization group approach. In
practice, this amounts to have the effective coupling
\be
\xi_{eff}\propto \ln ( Rl^2) \; ,
\ee
where $l^{-1}$ is a renormalization mass \cite{Ford87}.

To the best of our knowledge, no prescriptions for the value of 
$\xi$ other than those reviewed were proposed in the  
high energy physics literature. Instead, a strong prescription comes from
relativity.

In general relativity (and in all other metric theories of gravity in
which $\phi$ is a non-gravitational field\footnote{For example, the
Brans-Dicke scalar field is part of the gravitational sector of the
theory together with the metric tensor $g_{ab}$, hence it is a
{\em gravitational} scalar field.}), the only value of $\xi$ allowed by
the
Einstein equivalence principle \cite{Will} is the conformal coupling 
$1/6$. However, the derivation of this result \cite{SFFS} has
nothing to do with conformal invariance, conformal transformations, or
conformal flatness of the spacetime metric $g_{ab}$. It arises in the
study of wave propagation and tails of scalar radiation (violations of the
Huygens' principle) in curved spaces. This field of mathematical
physics is rather far from cosmology and, {\em a priori}, it is 
unlikely 
to contribute to cosmology, but this is not the case. Before getting into 
details, let us anticipate the main idea: one imposes that the
structure of tails of $\phi$ (which satisfies the wave equation
(\ref{KG})) becomes closer and closer to that occurring in Minkowski space
as the curved manifold is progressively approximated by its tangent
space. This is the Einstein equivalence principle \cite{Will} applied to
the physics of $\phi$ (of course, the rest of physics too has to satisfy
the Einstein equivalence principle; the requirement that $\phi$ does
satisfy it is only a necessary condition for consistency with general
relativity). 

Moreover, it turns out that $\xi=1/6$ is necessary in order to avoid
the physical pathology of {\em massive} fields $\phi$
propagating 
{\em along the light cones}.

We summarize now the derivation of this result: one begins with the  
physical definition of Huygens' principle due to Hadamard \cite{Hadamard}.
Assume that a point-like source of (scalar) radiation emits a delta-like
pulse at time $t=0$ in $r=0$. If at $t=t_1$ there is radiation only  on
the surface of the sphere with radius $r=ct_1$ and centre $r=0$, then we
say that Huygens' principle is satisfied (and that there are no tails). If
instead there is radiation also at radii $r<ct_1$ ({\em tails}), Huygens'
principle is violated.

Mathematically, the solution for a delta-like pulse is the retarded Green
function $G_R( x',x) $ of Eq.~(\ref{KG}), which satisfies

\be     \label{3s}
\left[ g^{a'b'}(x') \nabla_{a'}\nabla_{b'}-m^2-\xi R(x') \right]
G_R(x',x)=-\, \delta(x',x)   \; ,
\ee
where $\delta(x',x)$ is the four-dimensional Dirac delta  
\cite{DeWittBrehme} which satisfies the boundary condition 
$G_R(x',x)=0$ if 
$x'$ is in the past of $x$ and, for simplicity, we consider the case in
which the potential $V( \phi )$ reduces to a mass term (the
generalization to arbitrary potentials can be found in
Ref.~\cite{SFFS}). $G_R(x',x)$ has the general structure  
\cite{Hadamard,DeWittBrehme,Friedlander}
\be         \label{4s}
G_R(x',x)=\Sigma(x',x) \,\delta_R(\Gamma(x',x))+W(x',x)
\,\Theta_R(-\Gamma(x',x)) \; ,    
\ee
where $\Gamma(x',x)$ is the square of the geodesic distance 
between  
$x'$ and $x$ (a quantity well known in quantum field
theory in curved spaces \cite{BirrellDavies}); one has $\Gamma=0$ if $x'$
and $x$ are
light-like related, $\Gamma>0$ if $x'$ and $x$ are
space-like related, and $\Gamma < 0$ if $x'$ and $x$ are
time-like related. 
$\delta_R$  is the Dirac delta with support in the past of $x'$, 
and $\Theta_R$ is the Heaviside step 
function with support in the past light cone. The term in $\delta_R (
\Gamma) $ describes a contribution to the Green function from $\phi$ waves
propagating along the light cone ($\Gamma=0$), while the term  
$\Theta_R (- \Gamma) $ describes the contribution to $G_R$ from tails of
$\phi$ propagating {\em inside} the light cone ($\Gamma <0$). The
functions $\Sigma ( x', x )$ and $ W( x',x) $ are mere coefficients which
(at least in principle) are determined once the spacetime metric is fixed
\cite{DeWittBrehme,Friedlander}.

The Einstein equivalence principle is imposed as follows on the 
physics of the field $\phi$: when the spacetime manifold is
progressively approximated by its tangent space (i.e. by fixing 
the point $x$ and considering a small neighborhood of points $x'$
such that $x' \rightarrow x$), then the solution  
$G_R(x',x)$ for a delta-like pulse must reduce to the corresponding
one known from Minkowski spacetime, which is
\be   
G_R^{(Minkowski)}( x',x)=\frac{1}{4\pi} \, \delta_R( -\Gamma
)-\frac{m^2}{8\pi} \,  \Theta_R( \Gamma ) \; .
\ee
Calculations show that \cite{SFFS}
\be 
\lim_{x'\rightarrow x} \Sigma(x',x)=\frac{1}{4\pi} \;,
                       \label{5}
                              \ee
\be 
\lim_{x'\rightarrow x} W(x',x)=-\, \frac{1}{8\pi} \left[ m^2 + \left(
\xi-\frac{1}{6} \right) R(x) \right] \; ;
                       \label{6}
                              \ee
hence $G_R \rightarrow G_R^{(Minkowski)} $ if and only if
\be  \label{7}
\left( \xi-\frac{1}{6} \right) R(x) =0 \; ,
\ee
and this condition is verified, in general, only if $\xi=1/6$. Note that,
if
$\xi \neq 1/6$, a physical pathology may occur: the $\phi$-field can have
an arbitrarily large mass and still propagate along the light cone at the
spacetime points where Eq.~(\ref{7}) is satisfied; in this situation an
arbitrarily massive field would have no tails.   It is even possible to
construct an ``ultrapathological'' de Sitter spacetime in which the value
of the constant curvature and of the mass are adjusted in such a way that
a scalar field with arbitrarily large mass propagates along the light cone
at every point \cite{FaraoniGunzigIJMPD}.

The result that $\xi=1/6$ in general relativity is extended to all metric
theories of gravity in which $\phi$ is not part of the gravitational
sector \cite{SFFS}; in fact, in these theories, the Einstein
equivalence principle holds \cite{Will}; the fact that $\xi=1/6$ was 
confirmed in later studies \cite{GribPoberii}. 

\section{Inflation and $\xi \neq 0$: the unperturbed universe}

\setcounter{equation}{0}

In this section we summarize the consequences of the inclusion of NMC in
the
equations of the unperturbed Friedmann-Lemaitre-Robertson-Walker (FLRW)
universe. We assume that the metric is given by
\be  \label{metric}
ds^2=-dt^2+a^2(t) \left(  dx^2+dy^2+dz^2 \right) 
\ee
in comoving coordinates $\left( t,x,y,z\right)$.

It is clear from the previous section that one cannot arbitrarily set
$\xi=0$ and it was shown in several papers
\cite{Abbott81,FutamaseMaeda89,ALO90,FaraoniPRD96,Calgary}
that the value of $\xi$ determines the viability of inflationary
scenarios. The analysis of some specific inflationary scenarios was
performed in Ref.~\cite{FaraoniPRD96} and is not repeated here: it
suffices
to mention that a scenario should be examined with regard to:\\
{\em i)} theoretical consistence\\
{\em ii)} fine-tuning problems.

Regarding the former, one asks oneself whether any prescription for the
value of $\xi$ is applicable. If the answer is affirmative, one examines
the consequences for the viability of the specific scenario (does the
value of $\xi$ used correspond to the theoretical prescription~?). Aspects
studied include the existence of inflationary solutions and a sufficient
amount of inflation.

Fine-tuning is an aspect perhaps less fundamental but nevertheless  
important; the classic example is the already mentioned chaotic
inflationary scenario
with $V=\lambda \phi^4$ studied by Futamase and Maeda
\cite{FutamaseMaeda89}; inflationary solutions turn out to be fine-tuned
for $\xi \geq 10^{-3}$, in particular for the value $\xi=1/6\simeq 0.16$
predicted by general relativity.

\subsection{Necessary conditions for inflation}

In this section we study necessary conditions for inflation, defined as
acceleration of the scale factor, $\ddot{a} > 0$.
Acceleration of the universe, the essential qualitative feature of
inflation, is also required at the present epoch of the history of the
universe in order to explain the data from high redshift Type~Ia 
supernovae
\cite{SN}. The latter imply that a form of matter with
negative pressure (``quintessence'') is beginning to dominate the dynamics
of the universe. Scalar fields have been proposed as natural models of
quintessence \cite{Zlatevetal,Steinhardtetal,Chiba,Uzan,PBM},
and therefore,
the considerations of this subsection are also relevant for scalar
field models of
quintessence.

In ordinary inflation driven by a scalar field the
Einstein-Friedmann equations
\be  \label{dota} 
H^2 \equiv \left( \frac{\dot{a}}{a} \right)^2=\frac{\kappa}{3} \left(
\frac{\dot{\phi}^2}{2} + V( \phi) \right) \; ,
\ee  
\begin{equation}  \label{ddota}
\frac{\ddot{a}}{a}= - \, \frac{\kappa}{3} \left( {\dot{\phi}}^2 -V
\right)    \;  ,
\end{equation}
imply that a necessary (but
not sufficient) condition for cosmic acceleration is $V \geq 0$ (note
that in  slow-roll inflation 
$ \rho \simeq V( \phi ) >> \dot{\phi}^2 /2$ and in this case $V \geq 0$
is necessary to satisfy the weak energy condition \cite{Wald}). 

What is the analog of the necessary condition for inflation  when
$\xi \neq 0$~?  Manipulation of the equations of inflation with NMC
\cite{FaraoniPRD2000} yields
 \begin{equation}    \label{abc}
V-\frac{3\xi}{2}\phi \, \frac{dV}{d\phi} >0 \;\;\;\;\;\;\;\; \left( \xi
\leq 1/6 \right) \; .
\end{equation}
This necessary condition could not be generalized to values $\xi >1/6$,
due to the difficulty of handling the dynamical equations analytically
when $\xi \neq 0$ (no approximation was made). Albeit limited, the 
semi-infinite range of values of the coupling constant $\xi \leq 1/6$
covers many of the prescriptions for the value of $\xi$ given in the
literature.  In the $\xi \rightarrow 0 $ limit, Eq.~(\ref{abc}) reduces to
the well known necessary condition for acceleration $V>0$.

The necessary condition (\ref{abc}) immediately allows one to reach
certain conclusions:\\\\
{\em i)} consider an even potential $V( \phi )=V( -\phi )$
which
is increasing for $ \phi > 0$ (e.g. a pure mass term
$ m^2 \phi^2 /2$, a quartic potential, or 
their combination $V( \phi ) = m^2 \phi^2 /2 + \lambda \phi^4 +
\Lambda/\kappa $). For
$ 0 < \xi < 1/6$, one has $\xi \phi dV/d\phi >0 $ and it is harder to
satisfy the necessary
condition (\ref{abc}) for inflation than in the minimal coupling case.
Hence one can
say that, for this class of potentials, it is harder to achieve
acceleration of the universe, and hence inflation.
If instead $ \xi <0 $, the necessary condition for cosmic acceleration is
more
easily satisfied than in the $\xi=0 $ case,
but one is not entitled to say that with NMC it is easier to achieve
inflation (because a necessary, and not a sufficient
condition for acceleration is considered).\\\\
{\em ii)} Taking to the extremes the possibility of a balance between the
potential
$V( \phi)$ and the term $\xi R \phi^2/2$ in the
action (\ref{action}), one may wonder whether it is
possible to obtain inflation with a scalar field and $V( \phi )=0$ 
(i.e. a free, massless scalar with no cosmological constant, only owing to
the NMC. In particular, 
the case of  strong coupling $| \xi |
>> 1 $ considered many times in the literature
\cite{SalopekBondBardeen89,FakirUnruh90a,Morikawa,LTMI,BassettLiberati,
HwangNoh98,Chiba} is of interest.
It is immediate to conclude that this is not possible for $\xi \leq 1/6$
since by setting $V=0$ the necessary condition (\ref{abc}) cannot be
satisfied.

\subsection{The effective equation of state}

The effective equation of state 
\be
P=w\rho
\ee
of the cosmic fluid describing the scalar field has a coefficient $w$
that, in general, is time-dependent; it cannot be assigned {\em a priori}
without restricting the solutions to special ones ($a(t)= a_0 t^{2/
( 3( w+1))}$ if  $w\neq -1$, or $a=a_0$e$^{Ht}$ if $w=-1$) (solutions for
a
non-spatially flat universe and arbitrary values of $w$ can be found in
\cite{VFAmJP}). The function $w(t)$ depends on the particular solution of
the equations of motion.

In the case of minimal coupling and for a {\em general} potential $V$, the
effective equation of state of the universe is given by
\be
\frac{P}{\rho}=\frac{ \dot{\phi}^2 -2V}{ \dot{\phi}^2 +2V} \equiv w(x) \;
,
\ee
where $x \equiv \dot{\phi}^2 /2V $ is the ratio between the kinetic and
the
potential energy densities of the scalar $\phi$. Under the usual
assumption $V \geq 0$ (which guarantees that the energy density $\rho$
is non-negative when $\dot{\phi}=0$), one has that, for $x \geq 0$,
the function $w(x)=\left(x^2-1\right) \left( x^2+1 \right)^{-1} $
increases monotonically from its minimum $w_{min}=-1$
attained at $x=0$ to
the horizontal asymptote $+1$ as $x \rightarrow + \infty$. The slow
rollover 
regime corresponds to the region $|x | \ll 1$ and to $w$ near its minimum,
where the kinetic energy density of $\phi$ is negligible in comparison to
its potential energy density. As the kinetic energy density
$\dot{\phi}^2/2 $
increases, the equation of state progressively deviates from $P=-\rho$ and
the pressure becomes less and less negative; the system gradually moves
away from the slow rollover regime. At the equipartition between
the kinetic and the potential energy densities ($x=1$), one
has the ``dust'' equation of state $P=0$. The pressure becomes positive as 
$x$ increases and, when the kinetic energy density
completely dominates the potential energy density ($x \gg 1$), one
finally reaches 
the equation of state $P=\rho$.

The limitation $-1 \leq w(x) \leq 1 $ valid for $\xi =0$ does not hold for
$\xi \neq 0$: in the presence of NMC the ratio $P/\rho$ is not bounded
from below. An example is given by the exact solution with $P=-5\rho/3$
obtained in Ref.~\cite{IJTP2} for $\xi=1/6$ and corresponding to 
integrability of the equations of motion.

\section{Generalized slow-roll inflation}

\setcounter{equation}{0}

The equations or ordinary inflation are solved in the slow-roll
approximation; similarly, the equations for the density and gravitational
wave perturbations generated during inflation can only be solved, in
general,  in the same
approximation\footnote{The only exceptions are two specific
scenarios: power-law inflation and
the string-inspired scenario of Easther \cite{Easther}, a toy model
already ruled out by the {\em COBE} observations \cite{Easther}. Otherwise
one may resort to numerical integration in a specified scenario.}. Here,
the basics of the slow-roll approximation to ordinary inflation are
recalled, referring the
reader to the review paper \cite{Lidseyetal} and to
the references therein for a comprehensive discussion.

In the approximation
\be  \label{9}
\ddot{\phi} << H\dot{\phi} \; ,
\ee
\be  \label{10}
V( \phi ) \approx \rho >> \frac{\dot{\phi}^2}{2} \; , 
\ee
the equations of ordinary inflation (\ref{dota}), (\ref{ddota}) and
\be  \label{13}
\ddot{\phi}+3H\dot{\phi}+\frac{dV}{d\phi}=0 \; ,
\ee
simplify to 
\be  \label{14}
H^2\simeq \frac{\kappa}{3} V( \phi ) \; ,
\ee
\be  \label{15}
3H\dot{\phi}+\frac{dV}{d\phi} \simeq 0 \; .
\ee
In this approximation, the equation of state of the cosmic fluid
describing the scalar field is close to the vacuum equation of state
$P=-\rho$, and the cosmic expansion is almost a de Sitter one,
\be  \label{16}
a=a_0 \exp \left[ H(t) \, t\, \right] \; ,
\ee
with 
\be  \label{17}
H(t) = H_0 +H_1t+\ldots \; ,
\ee
where $H_0$ is a constant and dominates the (small) term $H_1t$ and the
next orders in the expansion (\ref{17}) of $H(t)$. The possibility that
the kinetic energy density $\dot{\phi}^2/2$ of the inflaton be negligible
in comparison with the potential energy density $V( \phi ) $ (as expressed
by Eq.~(\ref{10})) arises if $V( \phi ) $ has a flat section over which
$\phi$ can roll slowly (i.e. with small ``speed'' $\dot{\phi}$). This is a
necessary, but not sufficient, condition for slow-roll inflation to occur:
if $V( \phi ) $ is too steep, the inflaton will certainly roll fast down
the potential. Indeed, the slow-roll approximation is an assumption {\em
on the solutions} of the full equations of inflation (\ref{dota}),
(\ref{ddota}) and (\ref{13}). As a matter of fact, the potential could
have a flat section and $\phi$ could still shoot across it with large
speed $\dot{\phi}$. Moreover, it was noted \cite{Liddlereview} that
the reduced equations of slow-roll inflation (\ref{14}) and (\ref{15})
have degree reduced by one in comparison with the full equations of
ordinary inflation (\ref{dota}), (\ref{ddota}) and (\ref{13}). Hence,
the solution is specified by the reduced set of two initial conditions
$\left( \phi (t_0), a(t_0) \right) $ instead of the full set of four
conditions  $\left( \phi (t_0), \dot{\phi}(t_0), a(t_0), \dot{a}(t_0)
\right) $,
with an apparent loss of generality of the solutions. Then, why does the
slow-roll approximation work~? How is it possible that solutions of
slow-roll inflation be general solutions~? (if they correspond to a set of
zero measure in the set of all initial conditions, they are fine-tuned and
clearly unphysical). The answer is that the de Sitter solutions
\be \label{18}
a=a_0\exp( H_0t ) \; , \;\;\;\; \phi=0  
\ee
 are {\em attractor points} for the orbits of the solutions in the phase
space \cite{LiddleParsonsBarrow,SalopekBond}. Therefore, the
quasi-exponential expansion (\ref{16}) is  a {\em
general} property. Were inflationary (de Sitter) attractors absent,
slow-roll inflation would be an empty theory without generic solutions, a
formalism describing a speculation that doesn't occur in the real 
world.

How does the  attractor mechanism transfer  to the case of generalized
inflation~?
Does the attractor property of de Sitter solutions survive when NMC is
included in the picture~? Is a flat section of the potential still a
necessary condition for slow-roll inflation~? Regarding the last question,
it is useful to keep in mind that (as has been known for a long time
\cite{Abbott81,FutamaseMaeda89,FaraoniPRD96}) the NMC term $\xi R \phi^2
/2$ in the action (\ref{action}) acts as an effective mass
term\footnote{Although the effect is like that of a mass, 
the interpretation of the constant curvature
as a mass term for the scalar field must not be taken literally
\cite{FaraoniCooperstock95}.}, spoiling
the flatness of the potentials that are known to be inflationary for
$\xi=0$. These considerations will be re-examined and made quantitative in
the following.

One begins the analysis by writing the equations of generalized inflation
as
\be  \label{19}
6\left[ 1 -\xi \left( 1- 6\xi \right) \kappa \phi^2
\right] \left( \dot{H} +2H^2 \right) 
-\kappa \left( 6\xi -1 \right) \dot{\phi}^2   
- 4 \kappa V  + 6\kappa \xi \phi V' = 0 \; ,
\ee
\begin{equation}  \label{3}
\frac{\kappa}{2}\,\dot{\phi}^2 + 6\xi\kappa H\phi\dot{\phi}
- 3H^2 \left( 1-\kappa \xi \phi^2 \right) + \kappa  V =0 \, ,
\end{equation}
\be  \label{4}
\ddot{\phi}+3H\dot{\phi}+\xi R \phi +V' =0 \; .
\ee
Eqs.~(\ref{19})-(\ref{4}) are derived by varying the action
(\ref{action}); Eq.~(\ref{19}) corresponds to the trace of the Einstein
equations, $R=-\kappa ( \rho -3P )$; Eq.~(\ref{3}) is the Hamiltonian
constraint $3H^2=\kappa\rho$, while Eq.~(\ref{4}) is the well-known
Klein-Gordon equation (\ref{KG}).

Note that, in the presence of NMC, the energy-momentum tensor of the
scalar
field, and consequently its energy density $\rho$ and pressure $P$ can be
identified in several possible inequivalent ways, corresponding to
different ways of writing the field equations (see
Ref.~\cite{FaraoniPRD2000} for a detailed discussion). The procedure that
we adopt is identified as a convenient one in \cite{FaraoniPRD2000}
because it is general (i.e. solutions are not lost by manipulation of the
field equations) and the stress-energy tensor of the scalar field is
covariantly conserved, which may not happen for other choices of $\rho$
and $P$ \cite{FaraoniPRD2000}. 

Explicitly, the energy density and pressure of $\phi$ (which we assume
to be the only source of
gravity during inflation) relative to a comoving observer of the FLRW
universe are given by
\be
\rho =  \frac{\dot{\phi}^2}{2} +3\xi H^2\phi^2 +6\xi
H \phi \dot{\phi} + V( \phi )  \; ,  \label{22}
\ee
\be
P= \frac{\dot{\phi}^2}{2} - V( \phi ) -\xi \left( 4H
\phi\dot{\phi} +2 \dot{\phi}^2 +2 \phi \ddot{ \phi}\right) 
 -\xi \left( 2 \dot{H} +3 H^2 \right) \phi^2
 \, .  \label{23}
\ee
As discussed in detail in Ref.~\cite{Gunzigetal2000}, only two equations
of the set (\ref{19})-(\ref{4}) are independent, and the system can be
reduced to a two-dimensional phase space manifold with variables $\left(
H, \phi
\right)$. It is then straightforward to verify that the solutions
\be  \label{24}
\left( H, \phi \right) = \left( H_0, \phi_0 \right) \; ,
\ee
with $H_0$ and $\phi_0$ constants, are all the fixed points of the
dynamical system with $\xi \neq 0$, provided that the conditions 
\be  \label{C1}
12 \xi H_0^2 \phi_0 +V_0' =0 \; ,
\ee
\be  \label{C2}
H_0^2 \left( 1-\kappa \xi \phi_0^2 \right) = \frac{\kappa V_0}{3}  \; ,
\ee
are satisfied, where $V_0 \equiv V( \phi_0 )$ and $V'_0 \equiv \left. 
dV/d\phi \right|_{\phi_0}$. There are only two such constraints since
only two equations in the set
(\ref{19})-(\ref{4}) are independent.
The fixed points (\ref{24}) are de Sitter solutions with constant scalar
field and generalize the solutions $\left( H, \phi \right)=\left( \pm
\sqrt{ \Lambda/3}, 0 \right)$ well known for minimal coupling, $\Lambda >0
$ being the cosmological constant (corresponding to the constant potential
$V
=\Lambda/\kappa$).

In order to assess the stability of the
universes (\ref{24}) (i.e. to decide whether they are attractors or not), 
one has to perform a stability analysis with respect to
perturbations $\delta \phi$ and $\delta H$ of the scalar field and
the Hubble parameter\footnote{In the analysis of the phase space,
attention is usually restricted only to time-dependent perturbations
(e.g. \cite{Halliwell,Mulleretal}): however, these perturbations are too
special 
to draw definite conclusions.}, 
\be \label{25} \phi \left( t, \vec{x}
\right) =\phi_0 (t) + \delta \phi \left( t, \vec{x} \right) \;
, \;\;\;\;\;\;\;\;\;\;\; 
H \left( t, \vec{x} \right) = H_0 + \delta H \left( t, \vec{x} \right)  \;
.
\ee
Since the general perturbations are space-and time- dependent, one is 
faced with the recurrent problem of gauge-dependence in cosmology:
if the perturbation analysis is performed in a particular
gauge (of which many appear in the literature), one can never be sure that
the growing (decaying) modes are genuine perturbations 
and not pure gauge modes which can be removed by coordinate
transformations \cite{KT,Linde}. 

To solve the problem, one needs to perform a gauge-independent 
analysis: we adopt  the covariant and
gauge-invariant formalism of Bardeen \cite{Bardeen}, in the modified
formulation of Ellis, Bruni and Hwang  
\cite{EllisBruni,HwangVishniac,MukhanovFeldmanBrandenberger}.
We first present and discuss the results \cite{Faraoni2000PLA}, postponing
their derivation to
the final part of this section. For {\em expanding} de Sitter spaces
(\ref{24}) with $H_0>0$, there is {\em stability} (and therefore
(\ref{24}) is an attractor point) if
\be   \label{26}
V_0'' \geq \frac{V'_0}{\phi_0} \,   \frac{1-3\xi
\kappa\phi_0^2}{1-\xi\kappa\phi_0^2} \;\;\;\;\;\;\; ( \phi_0 \neq 0 ) \; ,
\ee

\be   \label{27}
V_0'' + 4 \xi \kappa V_0 \geq 0  \;\;\;\;\;\;\; ( \phi_0 =0 ) \; .
\ee
By contrast, the {\em contracting} fixed points (\ref{24}) with $H_0<0$
are
always unstable, like in the case of minimal coupling.

Stability depends not only on the form of the scalar field potential,
which is expected, but also on the value of $\xi$. It is only in
particular situations that the $\xi $-dependence disappears and stability
holds
irrespective of the value of $\xi$. This happens, for example: \\
{\em i)} if $V( \phi )$ has a minimum ($V'_0=0$ and $V_0''>0$) at
$\phi_0$; \\
{\em ii)} if $V= \Lambda/\kappa + \lambda \phi^n$ 
(including the case of a simple mass term $m^2\phi^2/2$) with $\Lambda,
\lambda \geq
0$. This space is stable for $n \geq 1+f(x)$,
where
\be  \label{28}
x \equiv \kappa\xi\phi_0^2 \; , \;\;\;\;\;\;\;\;\;\;
f(x)=\frac{1-3x}{1-x} < 1 \; .
\ee
The stability conditions (\ref{26}) and (\ref{27}) are deduced by assuming
that $0<x<1$; if $x>1$ a negative effective gravitational coupling 
$G_{eff} \equiv G \left( 1-\kappa\xi\phi_0^2 \right)^{-1} $ arises
\cite{FutamaseMaeda89,FaraoniPRD2000}. Furthermore,   
the slow-roll parameter $\epsilon_3$ defined in the  next subsection 
diverges if the unperturbed 
solution $\phi (t)$ crosses one of the critical values 
\be  \label{29}
\pm \phi_1 \equiv \pm \frac{1}{ \sqrt{\kappa\xi}} 
\ee
(which are defined for $\xi>0$), while the slow-roll parameter
$\epsilon_4$ diverges if $\phi( t )$ crosses one of the other critical
values  
\be \label{30}
\pm \phi_2 \equiv \pm
\frac{1}{\sqrt{\kappa\xi \left( 1-6\xi \right) }}
\ee
(which exist for $0< \xi < 1/6$).

Under the  usual assumption that $V $ be non-negative, the
Hamiltonian constraint
(\ref{3}) forces $| \phi|$ to be smaller than $\phi_2 $
\cite{FutamaseMaeda89,ALO90}; we further assume that $| \phi | < \phi_1$.
If instead $ \left| \phi  \right| > \phi_1$, the direction of the
inequality (\ref{26}) is reversed.

The case $\phi = \pm \phi_1 $ not considered so far corresponds to a class
of solutions with constant Ricci curvature containing a de Sitter
representative \cite{Gunzigetal2000}. However,  the latter is clearly
fine-tuned and
unstable with respect to perturbations $\Delta \phi$.

For $\xi=0$, Eq.~(\ref{C2}) yields $V'_0=0$ for the fixed point, while the
stability condition (\ref{26}) gives $V''_0 >0$; this happens, e.g., when
$V( \phi ) $ has  a minimum $\Lambda/\kappa$ in $\phi_0$, which
intuitively
corresponds to stability. A solution starting at any value of $\phi$ is
attracted towards the minimum; if $\phi$  identically coincides with
$\phi_0$ and there is no kinetic energy ($\dot{\phi}=0$), Eqs.~(\ref{22})
and (\ref{23}) yield the energy density $\rho=3\xi H_0^2\phi_0^2 +V_0=-P$
and the vacuum equation of state (corresponding to de Sitter solutions)
holds. 

If instead $V_0''<0$ and the potential has a maximum $V_0=\Lambda/\kappa$
in $\phi_0$, a solution starting near $\phi_0$ will run away from it.

When $\xi \neq 0$ the interpretation of the stability conditions
(\ref{26}) and (\ref{27}) is complicated by the balance between $V( \phi )
$ and $\xi R \phi^2/2$ in the action (\ref{action}). Eqs.~(\ref{26}) and
(\ref{27}) make precise the previous qualitative  considerations on this
balance in
Refs.~\cite{Abbott81,FakirUnruh92apj,FutamaseMaeda89,FaraoniPRD96}.

As a conclusion, {\em slow-roll inflation only makes sense for $\xi \neq
0$ when the
conditions (\ref{26}) or (\ref{27}) are satisfied}. In this case, the
expanding de Sitter spaces (\ref{24}) satisfying Eqs.~(\ref{C1}) and
(\ref{C2}) are attractor points in the phase space. One must be cautious
and check that Eqs.~(\ref{26}) or (\ref{27}) are satisfied before applying
the slow-roll formalism presented in the next section. 
The importance of the inflationary attractors is made
clear once again by the example of the {\em contracting} spaces
(\ref{24}), for which 
the slow-roll approximation is {\em exact} (in the sense that the
slow-roll
parameters $\epsilon_i$ defined in the next section vanish exactly). 
However, this bears no
relationship with  the actual inflationary solutions because the
contracting spaces (\ref{24}) are not attractors.

\subsection{Derivation of the stability conditions}

The derivation of Eqs.~(\ref{26}) and (\ref{27}) proceeds as follows: the
metric perturbations are identified by the quantities $A, B, H_L$ and
$H_T$ in the expression of the spacetime metric 
\be  \label{31}
ds^2=a^2(t) \left\{ -( 1+2AY) dt^2 -2BY_i dtdx^i +\left[ \delta_{ij}
(1+2H_L ) +2 H_T Y_{ij} \right] dx^i dx^j \right\} \; ,
\ee
where the $Y$ are scalar harmonics satisfying
\be  \label{32}
\nabla^2 Y = \left( \frac{ \partial^2}{\partial x^2}+\frac{
\partial^2}{\partial y^2}+
\frac{ \partial^2}{\partial z^2} \right) Y=-k^2Y \; ,
\ee
$Y_i$ and $Y_{ij}$ are related to the derivatives of the $Y$ by 
\be  
Y_i=\frac{1}{k^2} \partial_i Y \; ,
\ee
\be  \label{33}
Y_{ij}=\frac{1}{k^2} \partial_i \partial_j Y  +\frac{1}{3} \delta_{ij} Y 
\; , \ee
respectively \cite{Bardeen}, and $k$ is the eigenvalue defined by
Eq.~(\ref{32}). We shall use Bardeen's gauge-invariant potentials 
\be
 \Phi_H = H_L + \frac{\dot{a}}{k}\,
\left( B- \frac{a}{k}
\, \dot{H}_T \right)  \;,
\ee
\be   
 \Phi_A = A + \frac{\dot{a}}{k}\,
\left( B- \frac{a}{k}
\, \dot{H}_T \right) +\frac{a}{k} \left[ \dot{B}-\frac{1}{k} \left( a
\dot{H}_T \right)^{\dot{}} \right] \;,
\ee
and the Ellis-Bruni-Hwang 
\cite{EllisBruni} variables
\be   \label{DeltaphiR}
\Delta \phi \left( t, \vec{x} \right) = \delta \phi + \frac{a}{k}\,
\dot{\phi} \left( B
- \frac{a}{k}
\, \dot{H}_T \right) \; , \;\;\;\;\;\;\;\;\;\;\;\;\;\;\;\;\;
\Delta R \left( t, \vec{x} \right) = \delta R + \frac{a}{k} \,\dot{R}
\left( B - \frac{a}{k} \, \dot{H}_T \right) \; .
\ee
The evolution equations for the gauge-invariant  
variables $\Phi_{H,A}$ and $\Delta \phi$ were derived in
Ref.~\cite{Hwang90CQG}:
\be  \label{sr8}
\dot{\Phi}_H+\left( \frac{\xi \kappa \phi \dot{\phi}}{1-\kappa\xi\phi^2}
-H \right) \Phi_A 
-\frac{\kappa}{1-\kappa\xi\phi^2} \left \{ \xi \phi \Delta
\dot{\phi} +\left[ \xi \phi \left( \frac{ \dot{\phi}}{\phi} -H \right)
-\frac{\dot{\phi}}{2} \right] \Delta \phi \right\} =0 \; ,
\ee
\begin{eqnarray}  \label{sr9}
& & \left( \frac{k}{a} \right)^2 \Phi_H +\frac{1}{1-\kappa\xi\phi^2}
\left( \frac{3\xi^2\kappa\phi^2}{1-\kappa\xi\phi^2} + \frac{1}{2} \right)
\kappa \dot{\phi}^2 \, \Phi_A 
-\frac{1}{1-\kappa\xi \phi^2} 
\left\{ \left( \frac{3\xi^2\kappa \phi^2}{1-\kappa\xi\phi^2} + \frac{1}{2}
\right) \kappa \dot{\phi} \Delta \dot{\phi} \right. \nonumber \\
& & \left. +\left[ \left(  \frac{k}{a} \right)^2 \xi \phi 
-\ddot{\phi} 
\left( \frac{3\xi^2\kappa\phi^2}{1-\kappa\xi\phi^2} + \frac{1}{2} \right)
\right] \kappa \Delta \phi \right\} =0
\; ,
\end{eqnarray}
\be   \label{sr10}
\Phi_A + \Phi_H -\frac{2 \xi \kappa \phi \Delta \phi}{1-\kappa \xi \phi^2}
=0 \; ,
\ee
\begin{eqnarray}  \label{sr11}
& & \ddot{\Phi}_H +H \dot{\Phi}_H + \left( H- \frac{\xi \kappa \phi
\dot{\phi}}{1-\kappa\xi\phi^2} \right) \left( 2\dot{\Phi}_H -\dot{\Phi}_A
\right) -\frac{\kappa V }{1-\kappa \xi\phi^2} \, \Phi_A \nonumber \\ 
& & +\frac{\kappa}{1-\kappa\xi\phi^2} \left\{
- \xi\phi \Delta \ddot{\phi} +\left[ \frac{\dot{\phi}}{2} -2\xi \left(
\dot{\phi}+H\phi \right) \right] \Delta \dot{\phi} \right. \nonumber \\
& & \left. + \left[ \xi\phi \left( \kappa p_H-\frac{\ddot{\phi}}{\phi}
-\frac{2H\dot{\phi}}{\phi} \right) -\frac{V'}{2\kappa} \right]
\kappa \Delta \phi  \right\}=0 \; ,
\end{eqnarray}
\be  \label{sr12}
\Delta \ddot{\phi} + 3H \Delta \dot{\phi} + \left( \frac{k^2}{a^2}
 + \xi R +V'' \right) \Delta \phi + \dot{\phi} \left( 3
\dot{\Phi}_H -\dot{\Phi}_A \right) +2 \left( V' +\xi R \phi \right) \Phi_A
+ \xi \phi \Delta R =0 \; ,
\ee
where 
\be  \label{pressure}
p_H=\frac{1}{1-\kappa\xi\phi^2} \left[  \frac{\dot{\phi}^2}{2} -V -2\xi
\phi
\left( \ddot{\phi}+3H\dot{\phi} + \frac{\dot{\phi}^2}{\phi} \right)
\right]
\; .
\ee
An overdot denotes differentiation with respect to the comoving time of
the unperturbed background, and the
subscript zero denotes unperturbed
quantities. The formulation of Ref. \cite{Hwang90CQG}
has been adapted to the case of a FLRW universe with flat spatial
sections; the constant $\kappa=8\pi G $  has been restored. Only first
order calculations in the perturbations are presented here.

Considerable simplifications occur in Eqs.~(\ref{sr8})-(\ref{pressure})
for the case of a de Sitter space with constant scalar field (\ref{24}) as
the background universe;  to first order, one obtains
\be   \label{sr14}
\Phi_H=\Phi_A=\frac{\xi \kappa \phi_0}{1-\kappa\xi\phi_0^2} \, \Delta \phi
\; ,
\ee
\be  \label{sr15}
\Delta \ddot{\phi} + 3H_0 \Delta \dot{\phi} + \left[  
 \frac{k^2}{a^2}   +V_0''+
\frac{\xi R_0 \left( 1+\kappa\xi\phi_0^2 \right) +2V_0'
\kappa\xi\phi_0}{1-\kappa\xi\phi_0^2}  \right] \Delta \phi + 
\xi \phi_0 \Delta R =0 \; , 
\ee
while Eq.~(\ref{sr11}) reduces to the  constraint 
\be  \label{47}
-\, \frac{V_0\xi\phi_0}{1-\kappa\xi\phi_0^2}=V_0'-p_H\xi\phi_0 
\ee
which, using Eq.~(\ref{pressure}), is written as
\be  \label{48}
\frac{V_0'}{V_0}=-\, \frac{4\kappa\xi\phi_0}{1-\kappa\xi\phi_0^2} \; ;
\ee
Eq.~(\ref{48}) can also be obtained by division of Eq.~(\ref{C1}) by
Eq.~(\ref{C2}).
Using the fact that $R=6 ( \dot{H}+2H^2)$ and Eq.~(\ref{C1}) one obtains
\be  \label{49}
\Delta \ddot{\phi}+3H_0\Delta\dot{\phi}+\left( V_0''+12\xi H_0^2
+\frac{k^2}{a^2} \right) \Delta\phi +\xi\phi_0 \Delta R=0 \; .
\ee
For a de Sitter background (\ref{24}) the gauge-invariant variables
$\Delta \phi$ and
$\Delta R$ coincide, respectively, 
with the scalar field and curvature perturbations $\delta \phi $ and
$\delta R$, to first order,
\be  \label{50}
\Delta \phi=\delta\phi \; , \;\;\;\;\;\;\;
\Delta R =\delta R=6\left( \delta \dot{H} +4H_0\delta H \right) \; ,
\ee
and therefore\footnote{This expression agrees with the one following 
Eq.~(38) of Ref.~\cite{Hwang90CQG}.} 
\be    \label{51}
\Delta R =\delta R= \frac{-6\xi \kappa\phi_0 \left[ V_0''+4\left( 1+3\xi
\right) H_0^2 \right]}{1-\xi \left( 1-6\xi \right) \kappa\phi_0^2} \,
\Delta \phi \; .
\ee
One can then substitute Eq.~(\ref{51}) into Eq.~(\ref{49}) for
$\Delta\phi$ and use Eq.~(\ref{C1}) to obtain
\be  \label{sr17}
\Delta \ddot{\phi} + 3H_0 \Delta \dot{\phi} + 
\left( \frac{k^2}{a^2}   + \alpha \right) \Delta \phi  =0 \; , 
\ee
where 
\be  \label{alpha}
\alpha = \frac{V_0''\phi_0 \left( 1-\kappa\xi\phi_0^2 \right) -V_0' \left(
1-3\kappa \xi\phi_0^2 \right)}{\phi_0 \left[ 1-\xi \left( 1-6\xi
\right) \kappa \phi_0^2 \right]} 
\ee
and $a=a_0 \exp( H_0 t ) $. 
Let us consider the expanding ($H_0 >0$) de Sitter spaces (\ref{24}):
at late times $t \rightarrow + \infty$ one can neglect the 
$ \left( k/a \right)^2 \propto $e$^{-2H_0 t}$ term in Eq.~(\ref{sr17}) and
look for solutions of the form
\be    \label{sr19}
\Delta \phi \left( t, \vec{x} \right) = \frac{1}{(2\pi )^{3/2}} \int d^3
\vec{l} \,\, \Delta \phi_l (t) \, {\mbox e}^{i\, \vec{l} \cdot \vec{x} }
\; ,
\;\;\;\;\;\;\;   
\Delta \phi_l (t) =\epsilon_l \, \mbox{e}^{\beta_l \, t}   \, .
\ee
Note that the Fourier expansion (\ref{sr19}) is well defined because the
universe has flat spatial sections. 
The constants $\beta_l$ must satisfy the algebraic equation 
\be \label{56}
\beta_l^2+3H_0\beta_l +\alpha=0 \; ,
\ee
with roots
\be
\beta_l^{( \pm )} =\frac{3H_0}{2} \left(  -1 \pm \sqrt{
1-\frac{4\alpha}{9H_0^2}} \, \right) \; .
\ee
While {\em Re}$( \beta_l^{(-)} ) <0$, the sign of 
{\em Re}$ ( \beta_l^{(+)} ) $ depends on $\alpha$: 
{\em Re}$( \beta_l^{(+)} ) >0$ if $\alpha <0$ and
{\em Re}$( \beta_l^{(+)} ) \leq 0$ if $\alpha \geq 0$. Hence one  
has {\em stability} for $\alpha \geq 0$ which, for $\phi_0\neq 0$
translates into the advertised result (\ref{26}). If instead $\alpha <0$,
the gauge-invariant perturbations $\Delta\phi$ and $\Delta R \propto
\Delta\phi$ (cf.
Eq.~\ref{51})) grow without bound and there is {\em instability}.

Let us discuss now the $\phi_0=0$ case; 
Eqs.~(\ref{sr8})-(\ref{sr12}) yield
\be   \label{sr24}
\Phi_H=\Phi_A=0 \; ,
\ee
\be  
\Delta \ddot{\phi} +3H_0 \Delta \dot{\phi} + \left( \frac{k^2}{a^2}
 +\alpha_1 \right) \Delta \phi=0 \; ,
\ee
where $\Delta R =0$ and $\alpha_1= V_0''+4\xi \kappa V_0 $; hence, for
$\phi_0 =0$, there
is {\em stability} if Eq.~(\ref{27}) is satisfied and instability
otherwise.

Finally, consider the {\em contracting} ($H_0<0$) fixed points (\ref{24});
in
this case it is convenient to use conformal time $\eta$ (defined by $ dt
=ad\eta$) and the auxiliary variable $u \equiv a \Delta \phi$.
Eq.~(\ref{sr17}) becomes
\be   \label{sr29}
\frac{d^2u}{d\eta^2} + \left[ k^2 -U ( \eta ) \right] u=0
\; ,
\ee
where
\be  \label{sr30}
U ( \eta ) = \left( 4-\frac{\alpha_1}{H_0^2} \right)
\frac{1}{\eta^2}+\frac{2}{H_0 \eta^3} \; ,
\ee
and we used the relation 
\be  \label{62}
\eta =- \, \frac{1}{aH_0} 
\ee
valid in the background
(\ref{24}) (see Appendix~B). Formally, Eq.~(\ref{sr29}) is a
one-dimensional
Schr\"{o}dinger
equation for a quantum particle of unit mass in the potential $U( \eta )$;
its
asymptotic solutions  at large $\eta $ (corresponding to $t\rightarrow +
\infty$ for a contracting de Sitter background) are 
free waves $u \simeq {\mbox e}^{\pm i \, k \eta}$, and $\Delta \phi 
\propto H_0 \eta  $ diverges. The solutions (\ref{24}) with $H_0<0$ are
{\em unstable}, as in the $\xi=0 $ case.

\subsection{Slow-roll parameters}

The Hubble slow-roll approximation known for ordinary inflation
(\cite{Lidseyetal} and references therein) is characterized by two
slow-roll parameters $\epsilon_H=-\dot{H}/H^2$ and $\eta_H= -\ddot{\phi}/
( H \dot{\phi}) $ which stay small during slow-roll inflation. When
$\epsilon_H$ and $\eta_H$  increase, the kinetic energy of the inflaton
increases and, when $\epsilon_H$ and $\eta_H$ become of order unity, the
slow-roll approximation breaks down and inflation ends.

Slow-roll parameters have been identified also for generalized inflation
\cite{Hwang90CQG,Kaiser,Kaiserpreprint,HwangNoh96}; the novelty is that
there are four
such parameters as opposed to the two of ordinary inflation. From the
point of view of Sec.~3, this fact may provide a rationale of why
it is harder to achieve
slow-roll inflation with nonminimal rather than minimal coupling, for a
given potential $V( \phi )$: one has to satisfy four slow-roll
necessary conditions
instead of two. The slow-roll parameters are the dimensionless quantities
\be  \label{63}
\epsilon_1 = \frac{\dot{H}}{H^2}=-\epsilon_H \; , 
\ee
\be  \label{64}
\epsilon_2 =\frac{\ddot{\phi}}{H \dot{\phi}} =-\eta_H \; , 
\ee
\be  \label{65}
\epsilon_3 =-\, \frac{\xi \kappa \phi \dot{\phi}}{H \left[ 1- \left(
\frac{\phi}{\phi_1} \right)^2 \right]} \; , 
\ee
\be  \label{66}
\epsilon_4 =-\, \frac{\xi \left( 1-6\xi \right) \kappa \phi \dot{\phi}}
{H \left[ 1-\left( \frac{\phi}{\phi_2} \right)^2 \right] } \; .
\ee
$\epsilon_3$ and $\epsilon_4$  vanish in the limit $\xi \rightarrow 0 $ of  
ordinary inflation; $\epsilon_4$
also  vanishes for conformal coupling ($\xi=1/6$). One has 
$\left| \epsilon_i \right|
<< 1$ for every solution attracted by the expanding de
Sitter spaces (\ref{24}) (when the latter are attractor points) at
sufficiently large times.
Moreover, $\epsilon_i=0$ exactly for de Sitter solutions.

\section{Inflation and $\xi \neq 0$: perturbations}

\setcounter{equation}{0}

The quantum fluctuations of the inflaton field which unavoidably take
place during inflation generate density (scalar) perturbations that act as
seeds for the formation of the structures observed in the universe today,
from galaxies to superclusters \cite{KT,Linde}. Similarly, quantum
fluctuations
$\delta g_{ab}$ of the metric tensor  are generated during inflation,
corresponding to gravitational waves  
\cite{KT,Linde}. Both scalar and tensor perturbations leave an imprint on
the
cosmic microwave background by generating temperature fluctuations. The
latter have been detected by {\em COBE} \cite{smootetal} and other
experiments, and are going to be studied with unprecedented accuracy by
the
{\em MAP} \cite{MAP} and {\em PLANCK} \cite{PLANCK} satellites to be
launched in the very near future. 

In order to confront itself with the present and future observations, the
theory must predict observables such as the amplitudes and spectra of
perturbations. For ordinary inflation, these have been
computed\footnote{The calculation took a long time to be completed,
starting with the
early efforts of the early 80's (see Ref.~\cite{Lidseyetal} for a recent
review and Ref.~\cite{Guth} for a historical perspective).}. In a 
remarkable series of papers, Hwang and collaborators
\cite{Hwang} have performed a similar calculation for generalized
gravity theories described by the action
\be  \label{67}
S=\int d^4x \sqrt{-g} \left[ \frac{1}{2} f( \phi, R ) -\frac{\omega 
( \phi )}{2} \nabla^c \phi \nabla_c\phi -V( \phi ) \right] \; .
\ee
 The case of a nonminimally coupled scalar field is recovered by setting
\be  \label{68}
f( \phi, R) =\frac{R}{\kappa}-\xi R \phi^2 \; , \;\;\;\;\;\;
\omega =1 \; .
\ee
Hwang's treatment is covariant and gauge-invariant and builds upon the
formalism
developed by Bardeen \cite{Bardeen}, Ellis, Bruni and Hwang
\cite{EllisBruni} and Hwang and Vishniac \cite{HwangVishniac}, and
considers a FLRW universe with arbitrary curvature index. Motivated by
inflation, we restrict ourselves to the spatially flat case. The idea of
Refs.~\cite{Hwang} is to reduce the field equations of the  theory to
formal Einstein equations
\be   \label{64bis}
G_{ab}=\kappa T_{ab}^{(eff)} \; ,
\ee
where $T_{ab}^{(eff)} $ is an effective stress-energy tensor incorporating
terms that would normally appear in the left hand side of the field
equations. The treatment proceeds by using the gauge-invariant study of
perturbations
in Einstein gravity and ordinary inflation 
\cite{MukhanovFeldmanBrandenberger,Lidseyetal}.  

In the following, we review and complete the calculation, adapting it to
the case of the action (\ref{action}) and (\ref{68}); we believe that this
review is useful for
future reference, since a  consistent discussion of the slow-roll
approximation was not given before for generalized inflation. Instead of
using Eq.~(\ref{31}), it is
convenient to rewrite the metric perturbations in a different form.

\subsection{Scalar perturbations}

The metric is written as 
\be  \label{65bis}
ds^2=-(1+2\alpha)dt^2-\chi_{,i} dtdx^i +a^2(t) \delta_{ij} (1+2\varphi )
dx^idx^j \; ,
\ee
while the scalar field is given by Eq.~(\ref{25}). One introduces the
additional gauge-invariant variable 
\be  \label{66bis}
\delta \phi_{\varphi}=\delta \phi -\frac{\dot{\phi}}{H} \varphi \equiv -\,
\frac{\dot{\phi}}{H} \varphi_{\delta\phi} \; .
\ee
The second order action for the perturbations (analogous to the one for
ordinary inflation \cite{Lidseyetal}) is \cite{HwangQ}
\be  \label{67bis}
S_{pert}=\int dt d^3\vec{x} {\cal L}_{pert}= \frac{1}{2} \int dtd^3
\vec{x}
a^3 Z \left\{
\delta \dot{\phi}_{\varphi}^2 -\frac{1}{a^2} {\delta
\phi_{\varphi}}^{,i} {\delta \phi_{\varphi}}_{,i} +
\frac{1}{a^3 Z} \frac{H}{\dot{\phi}} \left[  a^3 Z \left(
\frac{\dot{\phi}}{H} \right)\dot{} \right] \dot{} \, 
\delta\phi_{\varphi}^2
\right\} \; ,
\ee
where 
\be  \label{68bis}
Z(t)=\frac{  H^2\left[ 1-\kappa\xi( 1-6\xi) \phi^2 \right]
(1-\kappa\xi\phi^2 )}{\left[  H (1-\kappa \xi \phi^2) -\xi\kappa\phi
\dot{\phi} \right]^2 } 
\ee
(cf. Eq.~(6) of Ref.~\cite{HwangQ} and use Eq.~(\ref{68})). The action
(\ref{67bis}) yields the evolution equation for the perturbations
$\delta\phi_{\varphi}$
\be  \label{68ter}
\delta \ddot{\phi}_{\varphi} +\frac{ (a^3 Z) \dot{} }{a^3 Z}\,
\delta\dot{\phi}_{\varphi} - \left\{ 
\frac{\nabla^2}{a^2} +\frac{1}{a^3Z} \frac{H}{\dot{\phi}} \left[ a^3 Z
\left( \frac{\dot{\phi}}{H} \right)\dot{} \right]\dot{} \right\}
\delta\phi_{\varphi}=0 \; .
\ee
By using the auxiliary variables\footnote{The variable $z$ of
Eq.~(\ref{69}) agrees with the $z$-variable of
Ref.~\cite{MukhanovFeldmanBrandenberger} and with the $z$ of
Ref.~\cite{Lidseyetal}
multiplied by the factor $\sqrt{Z}$ (note that $Z=1$ corresponds to
ordinary inflation).}
\be  \label{69}
z(t)=\frac{a\dot{\phi}}{H} \sqrt{Z} \; ,
\ee
\be  \label{70}
v(t, \vec{x} )=z \, \frac{H}{\dot{\phi}} \delta\phi_{\varphi} = a
\sqrt{Z} \delta\phi_{\varphi} \; , 
\ee
Eq.~(\ref{68ter}) is reduced to 
\be \label{71}
v_{\eta\eta}-\left( \nabla^2+\frac{z_{\eta\eta}}{z} \right) v=0 \; ,
\ee
where $\eta$ denotes conformal time. 

Quantization is achieved by assuming that the background is classical
while the perturbations have quantum nature. A Heisenberg picture is used
in which quantum operators change in time while the state vectors remain
constant (the vacuum state of the system
is identified with the adiabatic vacuum \cite{BirrellDavies}, and one
wants the vacuum state to
remain unchanged in time). The fluctuations $\delta \phi ( t, \vec{x} )$
of the
scalar field are associated to a quantum operator $\delta \hat{\phi} ( t,
\vec{x} ) $; similarly $\varphi \rightarrow \hat{\varphi}$, and the
gauge-invariant 
variable (\ref{66bis}) is associated to the quantum operator
\be  \label{72}
\delta \hat{\phi}_{\varphi}=\delta \hat{\phi} -\frac{\dot{\phi}}{H}\hat{
\varphi} 
\ee
(the hats denote quantum operators). The unperturbed quantities are
regarded as classical.

Since the three-dimensional space is flat, it is meaningful to perform a
Fourier decomposition of the operator $\delta \hat{\phi}_{\varphi} $,
\be  \label{73}
\delta \hat{\phi}_{\varphi}=\frac{1}{( 2\pi)^{3/2}} \int d^3\vec{k} \left[
\hat{a}_{k} \delta\phi_{\varphi_k}(t) {\mbox e}^{i \vec{k}\cdot \vec{x}} +
\hat{a}_{k}^{\dagger} \delta\phi^{*}_{\varphi_k}(t) {\mbox e}^{-i
\vec{k}\cdot \vec{x}} \right] \; ,
\ee
where the annihilation and creation operators $\hat{a}_k$ and
$\hat{a}^{\dagger}_k $ satisfy the canonical commutation relations
\be  \label{74}
\left[ \hat{a}_k, \hat{a}_{k'} \right] =
\left[ \hat{a}_k^{\dagger}, \hat{a}_{k'}^{\dagger} \right] =0 \; , 
\ee
\be  \label{75}
\left[ \hat{a}_k, \hat{a}_{k'}^{\dagger} \right] =
\delta^{(3)} \left(  \vec{k}-\vec{k}' \right)  \; , 
\ee
and the mode functions ${\delta \phi_{\varphi}}_{k} (t)$ are complex
Fourier coefficients satisfying the classical equations obtained from
Eq.~(\ref{68})
\be  \label{76}
{\delta \ddot{\phi}_{\varphi}}_k +\frac{ (a^3 Z)\dot{}}{a^3 Z}\,
{\delta\dot{\phi}_{\varphi}}_k + \left\{ 
\frac{k^2}{a^2} -\frac{1}{a^3Z} \frac{H}{\dot{\phi}} \left[ a^3 Z
\left( \frac{\dot{\phi}}{H} \right) \dot{} \right]  \right\}
{\delta\phi_{\varphi}}_k=0 \; .
\ee
Similarly,
\be  \label{77}
v( t , \vec{x} ) =\frac{1}{( 2\pi)^{3/2}} \int d^3\vec{k} \left[
v_{k} (t) {\mbox e}^{i \vec{k}\cdot \vec{x}} +
v_{k}^{*}(t) {\mbox e}^{-i
\vec{k}\cdot \vec{x}} \right] \; ,
\ee
\be  \label{78}
\hat{v}=\frac{zH}{\dot{\phi}} \delta \hat{\phi}_{\varphi} =a\sqrt{Z}
\, \delta \hat{\phi}_{\varphi} \; ,
\ee
and the $v_{k}(t)$ satisfy the equation
\be  \label{79}
(v_k)_{\eta\eta} + \left( k^2-\frac{z_{\eta\eta}}{z} \right) v_{k}=0 \; .
\ee
The momentum conjugated to $\delta \phi_{\varphi}$ is 
\be  \label{80}
\delta \pi_{\phi}( t, \vec{x} ) = \frac{\partial {\cal L}_{pert}}{\partial
( \delta \dot{\phi}_{\varphi} )} =a^3 Z \,  \delta \dot{\phi}_{\varphi}
(t,
\vec{x} ) \; ,
\ee
and the associated quantum operator is $\delta \hat{\pi}_{\varphi}$.
$\delta \hat{\phi}_{\varphi}$ and 
$\delta \hat{\pi}_{\varphi}$ satisfy the equal time commutation relations
\be  \label{81}
\left[ \delta \hat{\phi}_{\varphi}(t, \vec{x}), \delta
\hat{\phi}_{\varphi} (t, \vec{x}') \right] =
\left[ \delta \hat{\pi}_{\varphi}(t, \vec{x}), \delta
\hat{\pi}_{\varphi} (t, \vec{x}') \right] =0 \; ,
\ee
\be  \label{82}
\left[ \delta \hat{\phi}_{\varphi}(t, \vec{x} ), \delta
\hat{\pi}_{\varphi} (t, \vec{x}') \right] =\frac{i}{a^3 Z}
\delta^{(3)}\left( \vec{x}-\vec{x}' \right)  \; .
\ee
${\delta\phi_{\varphi}}_k (t) $  satisfy the Wronskian condition
\be  \label{83}
{\delta \phi_{\varphi}}_k  
{\delta \dot{\phi}_{\varphi}}_k^*-
{\delta \phi_{\varphi}}_k^* 
{\delta \dot{\phi}_{\varphi}}_k =\frac{i}{a^3 Z} \; .
\ee
In Refs.~\cite{Hwang} it is {\em assumed} that 
\be  \label{84}
\frac{z_{\eta\eta}}{z}=\frac{n}{\eta^2} \; ,
\ee
where $n$ is a constant; we will comment later on the validity of this
assumption. Under the assumption (\ref{84}), Eq.~(\ref{79}) for the
Fourier modes $v_k$ reduces to
\be  \label{85}
(v_{k})_{\eta\eta} + \left[ k^2 -\frac{( \nu^2 -1/4 )}{\eta^2} \right] v_k
=0 \; ,
\ee
where 
\be  \label{86}
\nu =\left( n+\frac{1}{4} \right)^{1/2} \; .
\ee
By making the substitutions
\be  
s=k\eta \; ,\;\;\;\;\; v_k=\sqrt{s}\,  J( s ) \; ,
\ee
Eq.~(\ref{85}) is reduced to the Bessel equation
\be
\frac{d^2 J}{ds^2}+\frac{1}{s} \frac{dJ}{ds} + \left( 1- \frac{\nu^2}{s^2}
\right) J=0 \; ;
\ee
therefore the solutions $v_k ( \eta )$ can be expressed as 
\be
v_k ( \eta) = \sqrt{k\eta} \, J_{\nu}( k\eta ) \; ,
\ee
where $J(_{\nu}( s) $ are Bessel functions of order $\nu$. Eq.~(\ref{70})
yields the solutions for the Fourier coefficients ${\delta
\phi_{\varphi}}_k$
\be  \label{87}
{\delta \phi_{\varphi}}_k( \eta ) =\frac{\dot{\phi}}{zH} v_k( \eta )
=\frac{1}{a\sqrt{Z}}v_k( \eta ) \; .
\ee
The $v_k( \eta) $ which solve the Bessel equation (\ref{85}) are expressed
in terms of Hankel functions $H^{(1,2)}_{\nu}$, leading to (see
Appendix~A)
\be   \label{88}
v_k ( \eta )=\frac{\sqrt{\pi |\eta |}}{2} \left[ c_1 ( \vec{k} )
H_{\nu}^{(1)}( k|\eta|) + c_2 ( \vec{k} ) H_{\nu}^{(2)} ( k|\eta| )
\right]
\ee
and, by using Eq.~(\ref{87}), to
\be  \label{89}
{\delta \phi_{\varphi}}_k ( \eta )= \frac{\sqrt{\pi|\eta|}}{2a\sqrt{Z}}
\left[ c_1 ( \vec{k} )
H_{\nu}^{(1)}( k|\eta |) + c_2 ( \vec{k} ) H_{\nu}^{(2)} ( k|\eta| )
\right] \; .
\ee
The normalization is chosen in such a way that the relation
\be  \label{90}
\left| c_2 ( \vec{k} ) \right|^2 -\left| c_1 ( \vec{k} ) \right|^2=1
\ee
holds, in order to preserve the equal time commutation relation
(\ref{83}). Furthermore, the coefficients are completely fixed by
requiring that, in the limit of small scales, the vacuum corresponds to
positive frequency solutions; in fact the field theory in Minkowski space
must be recovered in this limit\footnote{In other words, one is applying
again the Einstein equivalence principle \cite{Will}, this time to the
physics of quantum fluctuations of the field $\phi$.}. The small scale
(large wavenumber) limit corresponds to
\be  \label{91}
\frac{z_{\eta\eta}}{z} << k^2 \; ,
\ee
and Eq.~(\ref{79}) reduces to 
\be  \label{92}
(v_k)_{\eta\eta} -k^2v_k=0 
\ee
in this limit, with solutions $v_k \propto $e$^{\pm i k\eta}$. 
Eq.~(\ref{87}) yields 
\be  \label{93}
{\delta \phi_{\varphi}}_k=\frac{1}{a\sqrt{Z}\sqrt{2k}} \left[
c_1 ( \vec{k} )
{\mbox e}^{i k|\eta |} + c_2 ( \vec{k} ) {\mbox e}^{ -ik|\eta|}
\right] \; ,
\ee
which can also be obtained by expansion of the solutions (\ref{89}) for
$k|\eta | >>1$. Obviously, the positive frequency solution at small
scales is obtained by setting
\be
c_1 ( \vec{k} ) =0 \; , \;\;\;\;\;\; c_2( \vec{k})=1 
\ee
so that 
\be
\delta \phi ( \eta, \vec{x}) =\frac{1}{(2\pi )^{3/2}} \int d^3 \vec{x}
\left[ c_2 ( \vec{k}) {\mbox e}^{i( \vec{k}\cdot \vec{x} -k\eta )} +
c_2^* ( \vec{k}) {\mbox e}^{i( -\vec{k}\cdot \vec{x} +k|\eta | )} \right]
\; .
\ee

The {\em power spectrum} of a quantity $f( t, \vec{x} ) $ is defined as
\be  \label{95}
{\cal P}(k, t) \equiv \frac{k^3}{2\pi^2} \int d^3 r \, \langle f(
\vec{x}+\vec{r}, t) f( \vec{x},t) \rangle_{\vec{x}} \, {\mbox e}^{-i
\vec{k}\cdot\vec{r}} =\frac{k^3}{2\pi^2} \left| f_k (t) \right|^2 \; ,
\ee
where $\langle$~$\rangle_{\vec{x}}$ denotes an average over the spatial
coordinates $\vec{x}$ and $f_k(t)$ are the coefficients of the Fourier
expansion 
\be  \label{96}
f( t, \vec{x} ) =\frac{1}{(2\pi)^{3/2}}\int d^3 \vec{k} \, \left[ f_k (t)
\, {\mbox e}^{i \vec{k}\cdot \vec{x}} + 
f_k^* (t) \, {\mbox e}^{-i \vec{k}\cdot \vec{x}} \right] \; .
\ee
The power spectrum of the gauge-invariant operator $\delta
\hat{\phi}_{\varphi}$ is
\be  \label{97}
{\cal P}_{\delta\hat{\phi}_{\varphi}}(k, t) = \frac{k^3}{2\pi^2} \int
d^3 r \, \langle 0 | {\delta\hat{\phi}_{\varphi}}_k (
\vec{x}+\vec{r}, t) 
{\delta\hat{\phi}_{\varphi}}_k 
( \vec{x},t) |0 \rangle_{\vec{x}}\,\, {\mbox e}^{-i
\vec{k}\cdot\vec{r}} \; ,
\ee
where $\langle 0|\hat{A}|0\rangle$ denotes the expectation value of the
operator $\hat{A}$ on the
vacuum state. 
One is interested in computing the power spectrum for large-scale
perturbations, i.e. perturbations that cross outside the horizon during
inflation, subsequently remain ``frozen'' while outside the horizon, and
only after the end of inflation, during the radiation- or the
matter-dominated era re-enter the horizon to seed the formation of
structures. In the large scale limit the solution of Eq.~(\ref{68}) 
is
\be   \label{98}
\delta\phi_{\varphi}( t, \vec{x})= -\frac{\dot{\phi}}{H} \left[ C(
\vec{x}) -D( \vec{x}) \int^t dt' \, \frac{1}{a^3 Z}\, 
\frac{H^2}{\dot{\phi}}
\right] \; ,
\ee
where $C( \vec{x})$ and $D( \vec{x})$ are, respectively, the coefficients
of a growing component and of a decaying component that we neglect in the
following. Accordingly, the solution ${\delta \phi_{\varphi}}_k ( \eta )$
in the large scale limit $k|\eta | <<1 $ is
\be \label{99}
{\delta \phi_{\varphi}}_k ( \eta )= \frac{i\sqrt{|\eta |} \Gamma( \nu)}{2a
\sqrt{\pi Z}} \left( \frac{k|\eta |}{2} \right)^{-\nu} \left[ c_2 ( \eta)
-c_1 ( \eta ) \right] 
\ee
for $\nu \neq 0$, where $\Gamma$ denotes the gamma function. The power
spectrum (\ref{97}) therefore is given by
\be  \label{100}
{\cal P}^{1/2}_{\delta\hat{\phi}_{\varphi}}(k, \eta ) = 
\frac{\Gamma ( \nu )}{\pi^{3/2} a | \eta | \sqrt{Z}} \left(
\frac{k|\eta|}{2} \right)^{3/2-\nu} \left| c_2 ( \vec{k} ) -c_1 ( \vec{k}
) \right| 
\ee
for $\nu \neq 0$, while one obtains \cite{Hwang}
\be  \label{101}
{\cal P}^{1/2}_{\delta\hat{\phi}_{\varphi}}(k, \eta ) = 
\frac{2\sqrt{|\eta |}}{ a \sqrt{Z}} \left(
\frac{k}{2\pi} \right)^{3/2} \ln ( k | \eta | ) \left| c_2 ( \vec{k}) - 
c_1 ( \vec{k} ) \right| 
\ee
for $\nu=0$. Now, Eq.~(\ref{98}) yields (neglecting the decaying
component),
\be  \label{102}
C( \vec{x})=-\frac{H}{\dot{\phi}}\delta \phi_{\varphi}( t, \vec{x} ) 
\ee
and therefore, using Eq.~(\ref{95}),
\be  \label{103}
{\cal P}^{1/2}_C ( k, t ) = \left| \frac{H}{\dot{\phi}} \right| {\cal
P}^{1/2}_{\delta \phi_{\varphi}}( k, t ) \; .
\ee
By combining Eqs.~(\ref{66}) and (\ref{102}) one obtains
\be  \label{104}
\varphi_{\delta\phi}=-\, \frac{H}{\dot{\phi}} \delta \phi_{\varphi}=C \; .
\ee
The relation between temperature fluctuations of the cosmic microwave
background and the variable $C( \vec{x})$ is given in
Ref.~\cite{Hwang96PRD} as 
\be  \label{105}
\frac{\delta T}{T}=\frac{C}{5}
\ee
and therefore the spectrum of temperature fluctuations is
\be \label{106}
{\cal P}^{1/2}_{\delta T/T} (k, t)= \frac{1}{5}
{\cal P}^{1/2}_{C} (k, t)
\ee
and Eq.~(\ref{103}) yields
\be  \label{107}
{\cal P}^{1/2}_{\delta T/T} (k, t)= \frac{1}{5} \left|
\frac{H}{\dot{\phi}} \right| 
{\cal P}^{1/2}_{\delta \phi_{\varphi} } (k,
t)
\ee
with ${\cal P}^{1/2}_{\delta \phi_{\varphi}} $ given by Eqs.~(\ref{100})
and (\ref{101}). 

The {\em spectral index} of scalar perturbations is defined as
\be  \label{108}
n_s \equiv 1+ \frac{d\ln {\cal P}_{\delta \hat{\varphi}_{\delta\phi} 
}}{d\ln
k} \;, 
\ee
and Eq.~(\ref{106}) immediately yields
\be  \label{109}
n_s=1+\frac{d\ln {\cal P}_{C}}{d\ln k} \; .
\ee
By using Eqs.~(\ref{103}) and (\ref{100}) one obtains
\be  \label{110}
n_s=4-2\nu   \;\;\;\;\;\;\;\; ( \nu \neq 0) \; ;
\ee
in the following we do not need the expression for $\nu=0$. To proceed we
need extra input, which is connected to the validity of the assumption
(\ref{84}), which we now discuss. Hwang \cite{Hwang} proved that
Eq.~(\ref{84}) is satisfied for pole-like inflation $a(t) \propto
(t-t_0 )^{-q}$ ($q>0$), an expansion law appearing in pre-big bang
cosmology
related to low-energy string theory\footnote{Refs.~\cite{Hwang98CQG1}
and \cite{Hwang98CQG2} are
devoted, respectively, to the calculation of scalar and tensor
perturbations in pre-big bang cosmology.}. However, it turns out (a point
not discussed in these works) that {\em Eq.~(\ref{84}) is satisfied in
slow-roll inflation, to first order.} This is interesting for us because
we know, from Sec.~4, that for suitable values of $\xi$ there is a de
Sitter attractor (\ref{24}) for nonminimal coupling, and therefore that
it makes sense to consider the slow-roll approximation. 

Most models of ordinary inflation are set and solved in the context of
the slow-roll approximation. In generalized inflation as well, the
fact
that slow-roll conditions satisfy Eq.~(\ref{84}) allows one to solve
Eq.~(\ref{79}) for the perturbations.

Let us have  a deeper look at Eq.~(\ref{84}) and at the value of $n$ for
the
slow-roll approximation. The quantity $z_{\eta\eta}/z$ in Eq.~(\ref{79})
was computed exactly in Ref.~\cite{Hwang} in terms of the slow-roll
coefficients (\ref{63})-(\ref{66}). Upon assuming that
the $\epsilon_i$ be
small and that their derivatives $\dot{\epsilon}_i$ can be neglected
($i=1,...,4$), a situation mimicking the usual one \cite{Lidseyetal}, one
obtains to lowest order
\be  \label{112}
\frac{z_{\eta\eta}}{z}=a^2H^2 \left(
2-2\epsilon_1+3\epsilon_2-3\epsilon_3+3\epsilon_4 \right) \; .
\ee

The standard relation
\be  \label{113}
\eta\simeq -\, \frac{1}{aH} \, \frac{1}{1+\epsilon_1} \; ,
\ee
yields Eq.~(\ref{84}) with 
\be  \label{114}
n=2+3( -2\epsilon_1+\epsilon_2-\epsilon_3+\epsilon_4) 
\ee
and
\be  \label{115}
\nu=\frac{3}{2}- 2\epsilon_1 +\epsilon_2 -\epsilon_3 +\epsilon_4  \; .
\ee 
Eqs.~(\ref{110}) and (\ref{115}) yield the spectral index of scalar
perturbations for generalized slow-roll inflation,
\be  \label{116}
n_s=1+2 \left( 2\epsilon_1 -\epsilon_2 +\epsilon_3 -\epsilon_4
\right) \; ,
\ee 
where the right hand side is computed at the time when the perturbations
cross outside the horizon during inflation.  The deviations of $n_s$ from
unity (i.e. from an exactly scale-invariant Harrison-Zeldovich spectrum)
are small during generalized slow-rolling. For $\xi=0$ one recovers the
well
known formula for the spectral index  of ordinary inflation
\cite{Lidseyetal} $n_s=1-4\epsilon_H+2\eta_H$.

\subsection{Gravitational wave perturbations}

Tensor perturbations are generated as quantum fluctuations of the metric 
tensor $g_{ab}$ in nonminimally
coupled inflation, and  were calculated in Ref.~\cite{Hwang98CQG2} with a
procedure similar to the one used for scalar perturbations. We
briefly review also this calculation.

It is convenient to introduce tensor perturbations as the trace-free and
transverse quantities $c_{ij}$ in the metric
\be  \label{117}
ds^2=-dt^2+a^2(t) \left( \delta_{ij}+2c_{ij} \right) dx^i dx^j \; ,
\ee
with 
\be  \label{118}
{c^i}_i=0 \; , \;\;\;\;\;\;\; {c_{ij}}^{,j}=0 \; .
\ee
The power spectrum is
\be  \label{119}
{\cal P}_{c_{ij}}( k, t) =\frac{k^3}{2\pi^2} \int d^3 \vec{r} \, \langle
c_{ij}( \vec{x}+\vec{r}, t) c_{ij}( \vec{x}, t) \rangle_{\vec{x}} \, 
{\mbox
e}^{-i \vec{k}\cdot \vec{r}} 
\ee
and the spectral index of tensor perturbations is
\be  \label{120}
n_T= \frac{d\ln {\cal P}_{c_{ij}} }{d\ln k}  \; .
\ee
One obtains
\be  \label{121}
{\cal P}_{c_{ij}}^{1/2} ( k, \eta) =\frac{8\pi G H}{\sqrt{2}\pi} \frac{
1}{\sqrt{1-\xi
\kappa^2 \phi^2}} \frac{\Gamma ( \nu )}{\Gamma ( 3/2)} \left(
\frac{ k|\eta |}{2} \right)^{3/2-\nu_g} \sqrt{ \frac{1}{2} \Sigma_l \left|
c_{l1}( \vec{k}) -c_{l2}( \vec{k}) \right|^2 } \; ,
\ee
where the summation $\Sigma_l$ is intended over the two polarization
states $\times$ and $+$ of gravitational waves. In the slow-roll
approximation one has
\be  \label{122}
\frac{z_{\eta\eta}}{z}=\frac{m}{\eta^2} \; ,
\ee
where $m$ is the linear combination of the slow-roll parameters
\be  \label{123}
m=2-3( \epsilon_1-\epsilon_3) 
\ee
and $\nu_g=( m+1/4)^{1/2}$, as usual.  Hence, $\nu_g\simeq
3/2-\epsilon_1+\epsilon_3$ and, to first order,
\be  \label{124}
n_T=2 \left( \epsilon_1 - \epsilon_3 \right) \; .
\ee
Eq.~(\ref{124}) reduces to the well known spectral index of tensor 
perturbations \cite{Lidseyetal} of ordinary inflation when $\xi
\rightarrow 0$. Note that $n_T$ is very small for slow-roll inflation,
like in slow-roll ordinary inflation.

This completes the review of the calculation of spectral indices in
\cite{Hwang}; the reader is invited to consult the relevant
papers\footnote{A synthesis is given in Ref.~\cite{Kaiserpreprint} using
different variables.}. We just reported on the published work and
completed the calculation corresponding to the slow-roll regime. Slow-roll
inflation is not explicitly mentioned in Refs.~\cite{Hwang}, probably
because the attractor role of de Sitter solutions was not established at
that time. The knowledge that slow-roll inflation does indeed make sense
for $\xi \neq 0$ allows us to claim that Hwang's \cite{Hwang}
calculation applies to it, and to use Eq.~(\ref{116}) and (\ref{124}) for
the spectral indices to test generalized inflation with observations of
the cosmic microwave background.

\section{Open problems}

\setcounter{equation}{0}

The program of rethinking inflationary theory by including the (generally
unavoidable) NMC of the scalar field, a crucial ingredient too often
forgotten, is not exhausted by the results
presented in the previous sections.  In this section, we outline open
problems that constitute avenues for future research.

It may be useful to remark that, in addition to inflation, NMC changes the
description and the results of the dynamical systems approach to cosmology
(e.g. \cite{Gunzigetal2000}), quantum cosmology
\cite{QC,Okamura}, classical and quantum wormholes \cite{wormholes}, and 
constitutes a line of approach to the cosmological constant problem
\cite{Dolgov,Ford87,SuenWill}.

\subsection{Doppler peaks}

We did not present Doppler peaks for
generalized inflation: the
acoustic oscillations well known for minimal coupling
\cite{LiddleLyth} are indeed modified by NMC. Although a plot of the
Doppler peaks requires the specification of  a particular scenario of
generalized inflation (the potential $V( \phi ) $, the value of the
coupling constant $\xi$, details about the end of inflation and
reheating, etc), preliminary work was done in Ref.~\cite{PBM}.
 The acoustic peaks and the spectrum turnover are
displaced, and the effects of NMC in this model are \cite{PBM}:\\
{\em i)} an enhancement of the large scale, low multipoles $l$ region of
the Doppler peaks, due to an enhancement of the integrated Sachs-Wolfe
(or Rees-Sciama) effect;\\
{\em ii)} the oscillating region of the Doppler peaks is attenuated;\\
{\em iii)} the location of the peaks is shifted toward higher multipoles.

These features were derived by direct integration of the equations for the
perturbations with a modified {\em CMBFAST} code, under the assumption
that the
scalar field is the source of quintessence and that it has a potential of
the form $V( \phi ) \propto 1/\phi^{\alpha}$  ($\alpha >0$) in the
range of values spanned by $\phi$ today. 
Similar qualitative effects appear in a model based on induced gravity,
and
are interpreted as the signature of a broad class of scalar-tensor gravity
theories in the cosmic microwave background \cite{PBM}.

However, the nonminimally coupled scalar field driving inflation does not
necessarily have to be identified with the same scalar which possibly
constitutes quintessence today (as is instead done in quintessential
inflation). Such an identification, indeed, may appear artificial. If
the quintessence and the inflaton field do not coincide, the rather strong 
constraint \cite{Chiba,PBM} $\xi^2 G \phi_{today}^2 < 1/500$ coming from
tests of gravity in the Solar System  \cite{Will} is circumvented. More
important, the features of the Doppler peaks  could then be different. A
separate study is needed to explore their features in detail.

\subsection{Cosmic no-hair theorems}

Cosmic no-hair theorems \cite{KT} in the presence of NMC are not known:
one would like to know whether inflation is still a generic phenomenon
when $\xi \neq 0$. In other words, starting with an anisotropic Bianchi
model, does inflation occur and lead the universe toward a highly
spatially homogeneous and isotropic FLRW state, with flat spatial
sections~?

Preliminary results \cite{Starobinsky,FutRotMat89} show that
the convergence  to the $k=0$ FLRW universe can disappear  by going from
$\xi=0$ to $\xi \neq 0$.  Our perturbation analysis of Sec.~4 shows that,
when the deviations from homogeneity and isotropy are small, the de Sitter
solutions are still inflationary attractors in the phase space, but an
analysis for large deviations from a
FLRW space is needed.

\subsection{Reconstruction of the inflationary potential}

The reconstruction of the inflationary potential from cosmological
observations is  a task that was undertaken only for $\xi=0$. Given the
unavoidability of NMC in the general case, one would like to have a
similar formalism also for $\xi \neq 0$. Nothing has been done yet on
the subject. 

\subsection{A connection between inflation and boson stars~?}

Finally, we would like to point out a possible connection between
generalized
inflation and relativistic astrophysics. It appears from
observations of gravitational 
microlensing that there is  a population of objects with mass $M <
M_{\odot} $ responsible for the observed microlensing events. Boson stars
(see Ref.~\cite{Jetzer} for
a review) are candidates; nowadays, they are not regarded as 
unrealistic objects.

If boson stars exist at all, they are relics from the early universe 
since they are formed by bosons that were around at primordial times and
aggregated to form balls very early. Hence, there is the possibility
that boson stars are formed of a  condensate of inflaton particles. Since
the stability of boson stars depends on the value of $\xi$
\cite{VanderBij}, and such objects can only exists in a certain range of
values of $\xi$ \cite{Jetzer,VanderBij}, one can connect this range of
stability to successful generalized inflation scenarios (of which only a
few have been studied in the literature). This connection could be a link
between present-day objects and the early universe; work is in progress in
this direction. For example, gravitational lensing
by boson stars was studied for $\xi=0$ and shown
to have characteristic signatures \cite{DabrowskiSchunck}; this study
could be generalized to the $\xi \neq 0$ case.\\ \\

To conclude, the inclusion of NMC in the equations of inflation seems to
be necessary in most inflationary theories, and leads to important
consequences. Inflationary solutions are changed into noninflationary
ones, and fine-tuning problems appear. The much needed slow-roll
approximation to inflation is meaningful only when particular relations
between the scalar field potential, its derivatives, and the value of
$\xi$ are satisfied. In generalized slow-roll inflation, the spectra of
density and gravitational wave perturbations have beem computed, and are
given in Sec.~5. In our opinion, the most interesting problems left open
in generalized inflation are whether cosmic no-hair theorems hold, and the
reconstruction of the inflationary potential,  which
will be the subjects of future research.

\section*{Acknowledgments}

It is a pleasure to thank S. Sonego, M. Bruni, E. Gunzig, S. Odintsov, C.
Baccigalupi, V. Sahni, L. Ford, J. Miller, S. Liberati and E. Calzetta for
useful discussions.  This work was supported by the grants EC-DGXII
PSS*~0992 and HPHA-CT-2000-00015 and by OLAM, Fondation pour la Recherche
Fondamentale, Brussels. 

\clearpage

\section*{Appendix~A}

\def\theequation{A.\arabic{equation}}\setcounter{equation}{0}
\setcounter{equation}{0}

The Bessel function $J_{\nu}(s) $ can be expressed as 
\be
J_{\nu}(s) =\frac{H_{\nu}^{(1)}(s) +H_{\nu}^{(2)}(s)}{2} \; ,
\ee
and therefore
\be
v_k( \eta) = \sqrt{k\eta} J_{\nu}( k\eta )=
\frac{\sqrt{k\eta}}{2} \left[ H^{(1)}_{\nu}( k\eta )
+H^{(2)}_{\nu}( k\eta ) \right] \; .
\ee
In addition, the property 
\be
J_p(z)=\Sigma_{k=0}^{\infty} \frac{(-1)^k}{k! \Gamma( p+k+1)} \left(
\frac{z}{2} \right)^{p+2k} 
\ee
yields
\be
J_p( -z) =(-1)^p J_p(z) \; .
\ee
Then, $J_{\nu}( k\eta )=(-1)^{\nu} J_{\nu}(
k|\eta | )$ if $\eta <0$. 

\clearpage

\section*{Appendix~B}

\def\theequation{B.\arabic{equation}}\setcounter{equation}{0}
\setcounter{equation}{0}

In de Sitter space, the dependence of the scale factor on the comoving
time $t$
\be  \label{B1}
a=a_0\exp( H_0 t )
\ee
and the definition of conformal time
\be  \label{B2}
\eta=\int^t \frac{dt'}{a(t')} \; ,
\ee
yield the relation
\be  \label{B3}
\eta=-\, \frac{1}{aH_0}=-\frac{{\mbox e}^{-H_ot}}{a_0 H_0} \; .
\ee
For {\em expanding} de Sitter spaces ($H_0>0$), $t \rightarrow +\infty$
corresponds to $\eta \rightarrow 0$, while for {\em contracting} ($H_0<0$)
de Sitter spaces, $t \rightarrow +\infty$
corresponds to $\eta \rightarrow +\infty $. During slow-roll inflation,
Eq.~(\ref{B3}) is corrected to
\be  \label{B4}
\eta=-\, \frac{1}{aH} \, \frac{1}{1+\epsilon_1} \; ,
\ee
which holds in the approximation that the derivatives of the slow-roll
parameters $\epsilon_i$ can be neglected \cite{HwangQ}.

\end{document}